\documentclass{article}

\usepackage{lineno}
\usepackage{siunitx}
\usepackage{color,soul}
\usepackage{braket}
\usepackage{subcaption}
\usepackage{graphicx}
\usepackage{amsmath}
\usepackage{url}
\usepackage{amsfonts}

\newcommand{\inn}{\mathrm{in}}
\newcommand{\out}{\mathrm{out}}
\newcommand{\ah}{\hat{a}}
\newcommand{\ahd}{\hat{a}^\dagger}
\newcommand{\w}{\omega}
\newcommand{\dd}{\mathrm{d}}

\title{Phase-sensitive characterization of a quantum frequency converter by spectral interferometry}

\author{Mateusz J Olszewski$^{1*}$, Kasper Hecht Alexander$^{1,2}$, Michael T M Woodley$^1$,\\ Leah R Murphy$^1$, Peter J Mosley$^{1,3}$ and Alex O C Davis$^1$}

\begin{document}
\maketitle

\emph{$^1$Centre for Photonics, Department of Physics, University of Bath, BA2 7AY, United Kingdom}\\
\emph{$^2$Department of Electrical and Photonics Engineering, Technical University of Denmark, 2800 Kongens Lyngby, Denmark}
\\ \emph{$^3$ ORCA Computing Ltd., 30 Eastbourne Terrace, London W2 6LA, United Kingdom} \\
{$^*$mjo41@bath.ac.uk} \\



\centering
\subsection*{Abstract} 
We introduce an experimental technique for complete phase-sensitive characterization of arbitrary unitary spectral-temporal transformations of optical modes. Our method recovers the complex spectral transfer function, or Green's function, of a frequency converter by analyzing spectral interference in the response to a tunable bichromatic probe. We perform a proof-of-concept experiment on a frequency conversion module based on Bragg-scattering four-wave mixing in photonic crystal fiber. Our results validate our technique by recovering useful information in the phase of the Green's function, revealing the relative positions of regions of active frequency conversion and passive dispersive propagation within the module. Our work introduces a new approach to characterizing the performance of a variety of active devices with diverse applications in emerging quantum technologies.
\flushleft




\section{Introduction}


Quantum frequency conversion (QFC), where the spectrum of an optical signal is modified while its quantum state is preserved, is emerging as a crucial requirement in quantum technologies \cite{kimble2008quantum,walmsley2016building}. The applications of QFC include spectrally multiplexed single-photon sources \cite{joshi2018frequency}, hybridizing spectrally incompatible quantum devices \cite{bonsma}, and minimizing losses in quantum signals by optimizing their wavelength for transmission through fiber networks \cite{morrison2021bright,da2022pure,yu2023telecom} or for more efficient detection \cite{rakher2010quantum}. Recent demonstrations of city-scale quantum networks have employed QFC to enable long-distance entanglement distribution that can be mapped into quantum memories \cite{kucera2024demonstration,liu2024creation}, and QFC has likewise enabled  quantum state teleportation between photons emitted from remote quantum dots \cite{strobel2025telecom}. QFC has also been used to characterize ultrafast quantum pulses \cite{golestani:22,maclean2018direct}, violate optical reciprocity \cite{otterstrom2021nonreciprocal} and perform logical operations \cite{lu2020fully}. In pursuit of these applications, a variety of coherent processes have been exploited to implement QFC, including electro-optic modulation \cite{wright2017spectral}, $\chi^{(2)}$ processes such as sum/difference frequency generation \cite{PhysRevLett.68.2153,lio,zaske, morrison2021bright}, and $\chi^{(3)}$ processes such as cross-phase modulation \cite{matsuda2016deterministic} and Bragg-scattering four-wave mixing (BS-FWM) \cite{mcgarry2024microstructured}.  As the importance of QFC for future quantum technologies becomes clearer and the range of available methods expands, there is a growing demand for general techniques to characterize its performance. Typically, experimental demonstrations simply report the conversion efficiencies for a sample of input states, but this falls short of a full characterization and leaves important properties of the QFC process underdetermined (see next section). Previous work on ultrafast quantum pulse gates \cite{Eckstein:11} has achieved a full tomography of a QFC process based on sum-frequency generation within a finite mode basis \cite{PhysRevA.96.063817}. Similarly, an electro-optic frequency domain quantum processor has been fully characterized by probing it with a tomographically complete basis of inputs prepared using a pulse shaper \cite{Lu:23}. These techniques involve shaping the temporal mode of a probe pulse into a number of complex profiles that scales rapidly with the size of the mode basis. The speed and fidelity of the reconstruction are dependent on \emph{a priori} knowledge of the natural mode basis, or Schmidt decomposition, of the QFC, and reconstruction for highly multimode pulse gates requires a large number of measurements. 

In this work, we propose and experimentally demonstrate an alternative technique, which we name \emph{two-tone tomography}. Inspired by ultrafast pulse characterization techniques, this consists of probing the QFC with a bichromatic signal with small frequency separation, spectrally resolving the interference between the converted outputs of these two frequency components, and recovering the properties of the QFC with Fourier-domain analysis. This approach separately reconstructs the amplitude and the phase of the spectral transfer function, or Green's function, of the QFC. No \emph{a priori} knowledge of the QFC is required except the spectral span of the input and output. Two-tone tomography is well-suited to highly multimode QFC, is generalizable to all unitary time-frequency mode converters, and can be carried out using generally available components.

\section{The Green's function and significance of spectral phase}
\begin{figure}[h!]
    \centering
    \begin{subfigure}{0.49\textwidth}
        \includegraphics[width=\linewidth]{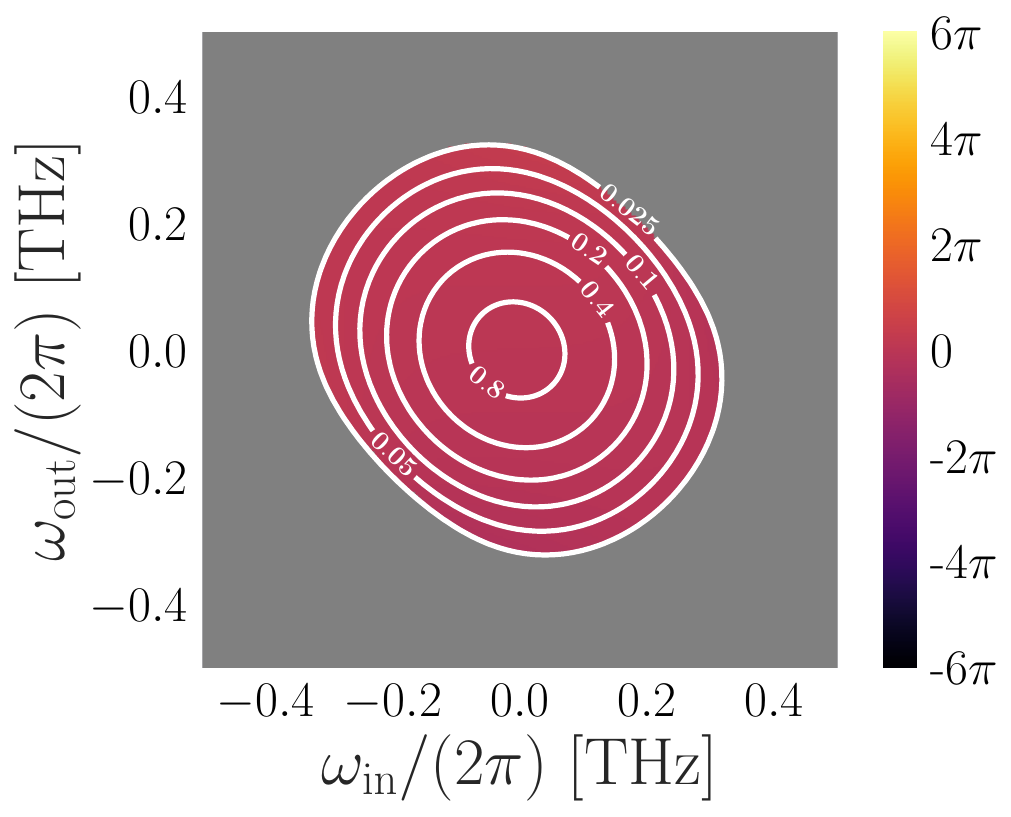}
        \caption{}
        \label{fig:Simulation_unchirped-pumps_GF}
    \end{subfigure}
    \hfill
    \begin{subfigure}{0.49\textwidth}
        \includegraphics[width=\linewidth]{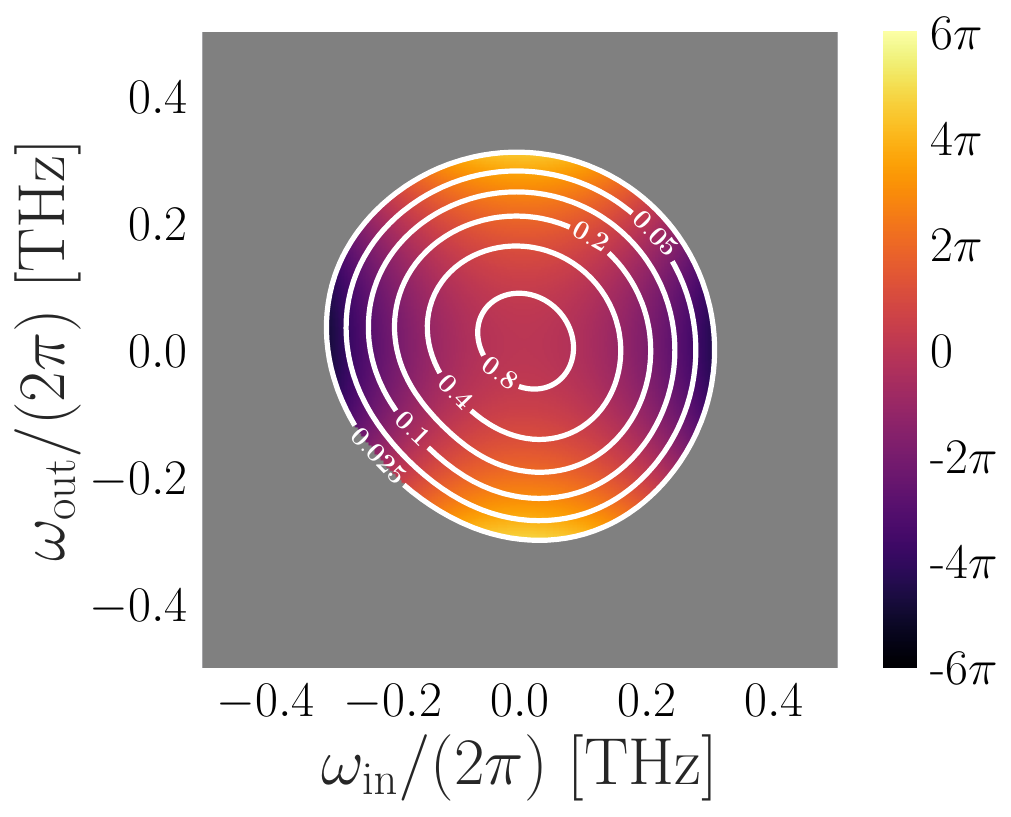}
        \caption{}
        \label{fig:Simulation_chirped-pumps_GF}
    \end{subfigure}
    \\
    \begin{subfigure}{0.49\textwidth}
        \includegraphics[width=\linewidth]{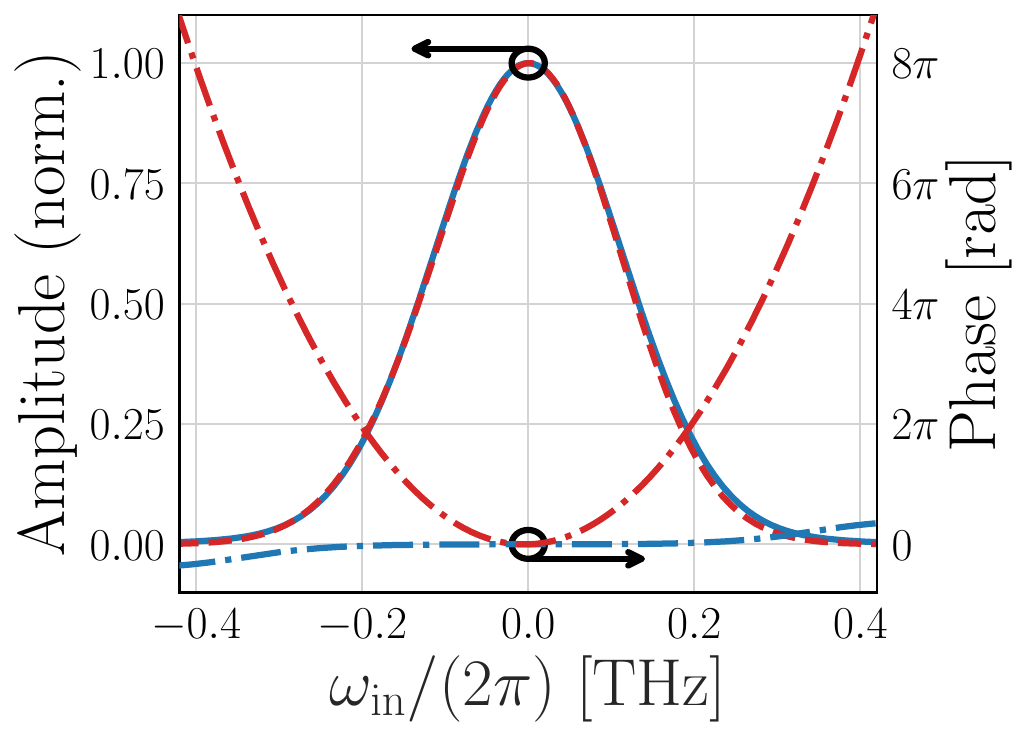}
        \caption{}
        \label{fig:Schmidt_modes_BOTH_spectral-domain}
    \end{subfigure}
    \hfill
    \begin{subfigure}{0.49\textwidth}
        \hspace{-1mm}\includegraphics[width=0.88\linewidth]{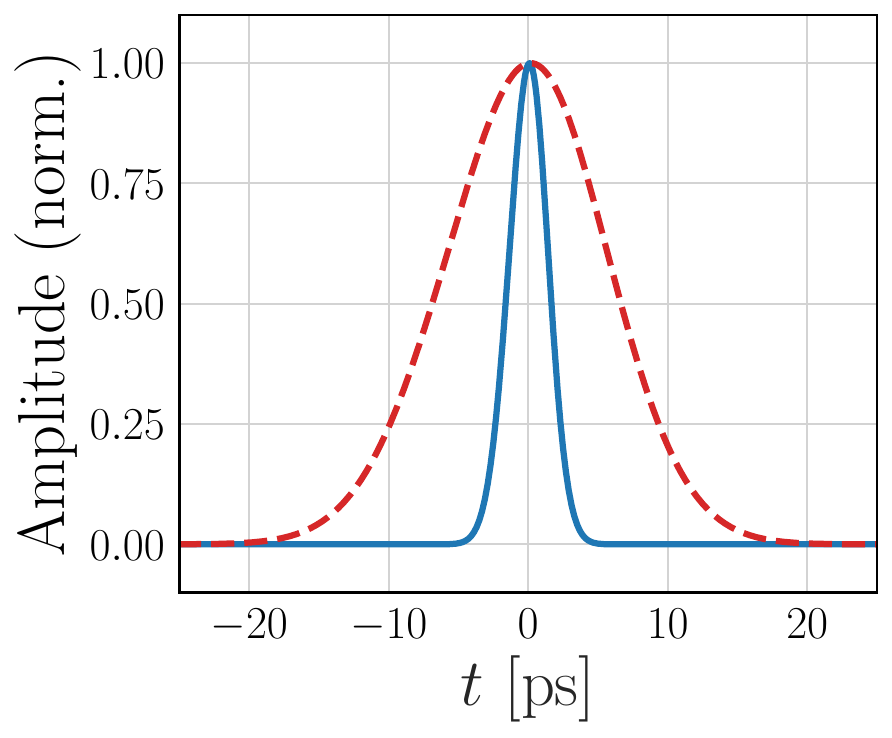}
        \caption{}
        \label{fig:Schmidt_modes_BOTH_time-domain}
    \end{subfigure}
    \caption{(a) The complex spectral transfer (Green's) function, $G(\w_\out,\w_\inn)$, for the case of unchirped pulsed pumps. The heatmap shows the phase, $\phi(\w_\out,\w_\inn)$, whereas the white concentric lines denote contours of the (normalized) magnitude, $|G(\w_\out,\w_\inn)|$. (b) Same as (a) but with chirped pumps. (c) The optimized input mode, $f(\omega_\inn)$, for maximum conversion in the two cases: unchirped pumps (blue), and chirped pumps (red). The spectral amplitude is shown with a solid line in the unchirped case and with a dashed line in the chirped case. The spectral phase is shown with dash-dotted lines in both cases. (d) The optimized input pulse in the time domain in the unchirped case (blue, solid) and chirped case (red, dashed), respectively.}
    \label{fig:SIM_results}
\end{figure}

 In single-photon QFC, where the pumps are considered to be undepleted, the process is linear in the input field and can be fully described by a Green's function $G(\omega_\out,\omega_\inn)$ \cite{Mejling:12} governing the input-output relation:
\begin{equation}
    \ah_{\out}(\omega_\out) = \int G(\omega_\out,\omega_\inn)\ah_\inn(\omega_\inn)\dd\omega_\inn,
    \label{Gdef}
\end{equation} 

\noindent where $\ah_{\mathrm{in}}(\w_\inn)$ and $\ah_{\mathrm{out}}(\w_\out)$ are respectively the annihilation operators for the monochromatic input and output modes with frequencies $\w_\inn$ and $\w_\out$. This model is general enough to account for arbitrary unitary transformations of the spectral-temporal mode, and can accommodate insertion loss in the postselected single-photon case. The absolute value $|G(\omega_\out,\omega_\inn)|^2$ is simply the intensity response of the QFC at frequency $\omega_\out$ to a monochromatic input at $\omega_\inn$, and can be recovered by probing the QFC with a tunable seed \cite{alexander2025characterizing}. However, the spectral phase $\phi(\omega_\out,\omega_\inn)=\arg \{G(\omega_\out,\omega_\inn) \} $ is equally important to understanding QFC with non-monochromatic inputs. For example, for a single-photon input in a state \cite{karpinski2021control}
\begin{equation}
    \ket{1_\inn}\equiv\int f(\w)\ahd(\w) \dd\w\ket{0},
\end{equation}
for some spectral mode $f(\w)$, the overall efficiency of QFC into some frequency band $\Delta\w$ is
\begin{equation}
\int_{\Delta\w}\braket{1_\inn|\hat{n}_\out(\w_\out)|1_\inn}\dd\w_\out=\int_{\Delta\w} \bigg|\int G(\w_\out,\w_\inn)f(\w_\inn) \dd\w_\inn\bigg|^2\dd\w_\out,
\label{efficiencyequation}
\end{equation}
where $\hat{n}_{\out}(\w_\out)=\ahd_\out(\w_\out)_\out\ah(\w_\out)$. In general, both this efficiency and the input pulse mode $f(\w_\inn)$ that optimizes it are strongly dependent on $\phi(\omega_\out,\omega_\inn)$.
An example of how the spectral phase determines whether an input pulse matches the temporal aperture of the interaction is shown in Fig.~\ref{fig:SIM_results}, which compares the Green's function for BS-FWM driven by unchirped and linearly chirped Gaussian pumps (simulated following Ref.~\cite{Korsgaard2024RamanScattering} -- see details in Supplemental Document). In the unchirped case (Fig.~\ref{fig:Simulation_unchirped-pumps_GF}), the phase of $G(\omega_\out,\omega_\inn)$ is nearly flat. Introducing a linear pump chirp leaves the Green's function amplitude almost unchanged (Fig.~\ref{fig:Simulation_chirped-pumps_GF}) but induces a quadratic spectral phase along the input- and output-frequency axes. Consequently, as seen in Fig.~\ref{fig:Schmidt_modes_BOTH_spectral-domain}, whereas the optimal input mode for the unchirped-pump case is simply a Gaussian with a nearly flat spectral phase matching the unchirped pumps (blue curves)~\cite{Mejling:12,PhysRevA.85.053829}, the optimal input mode for the chirped case acquires quadratic phase, although it retains the same Gaussian amplitude profile (red curves). In the temporal domain, Fig.~\ref{fig:Schmidt_modes_BOTH_time-domain}, the discrepancy in the spectral phase becomes a pronounced distortion of the optimized input pulse shape. Naively supplying an unchirped input pulse in the latter case leads to poor overlap with the temporal region in which the conversion occurs and results in a reduction in overall conversion efficiency exceeding 50\%. This illustrates that recovering the spectral phase, $\phi(\omega_\out,\omega_\inn)$, is essential for identifying the true optimal input mode and achieving maximum conversion efficiency. 

The phase $\phi(\omega_\out,\omega_\inn)$ can also encode important information about the internal dynamics of QFC, such as the longitudinal efficiency profile. In some platforms such as optical fiber, QFC may not be uniformly sustained along the length of the converter due to variations in the medium or pump fields. The phase of $G(\omega_\out,\omega_\inn)$ can reveal this in several ways. For example, if only one section of the medium is active for QFC, then the photon will undergo ordinary dispersive propagation prior to the region of active conversion. The length of this passive section, $L_\inn$, will then appear in $\phi(\omega_\out,\omega_\inn)$ as an additive contribution $L_\inn\phi_\inn'(\w_\inn)$, where $\phi_\inn'(\w_\inn)$ is the accumulated phase per unit length at frequency $\w_\inn$, which can be obtained from the dispersion profile of the device. This example illustrates how $\phi(\omega_\out,\omega_\inn)$ can lift the lid of the ``black box'' of QFC, such as by revealing the relative positions of regions of active nonlinear interaction and passive dispersive propagation within the device.

\section{Phase reconstruction by two-tone tomography}\label{theory}
$G(\omega_\out,\omega_\inn)$ is in several respects similar to the biphoton joint spectral amplitude (JSA) \cite{karpinski2021control}: both are two-dimensional complex functions relating the spectral-temporal mode of one photon to another, and both often derive from closely related parametric processes. In \cite{davis2020measuring}, the JSA was fully characterized by spectral shearing interferometry. Following an analogous approach, here we propose and experimentally demonstrate a technique to fully characterize $G(\omega_\out,\omega_\inn)$ interferometrically by probing the QFC with a bichromatic (``two-tone'') probe in a coherent state 
\begin{equation}
    \alpha_{\mathrm{in}}(\omega_\inn, \tau)=\frac{\alpha_{\mathrm{in}}^0}{\sqrt{2}}[e^{i\tau\Omega/2}\delta(\omega_0+\Omega/2-\w_\inn)+e^{-i\tau\Omega/2}\delta(\omega_0-\Omega/2-\w_\inn)]
\end{equation}
    with (tunable) central frequency $\w_0$, frequency spacing $\Omega$, beat delay $\tau$, and amplitude $\alpha^0_\inn$ (which we assume is real without loss of generality). In the time domain, this signal has intensity beats with period $2\pi/\Omega$. Unlike a constant-intensity monochromatic seed, this time-varying intensity allows the temporal response of the QFC to be probed, which enables the spectral phase information to be accessed. From Eq.~\eqref{Gdef} the spectral intensity at the output of the QFC is 
    \begin{align}
        I_\out(\w_\out,\tau)= \braket{\hat{n}_\out(\w_\out,\tau)}=\frac{|\alpha^0_\inn|^2}{2}\left(I_0(\w_\out)+\tilde{I}_\out(\w_\out)e^{i\Omega\tau}+\tilde{I}_\out^*(\w_\out)e^{-i\Omega\tau}\right),
        \label{eq:OutputIntensity}
    \end{align}

\noindent which includes the phase-dependent interference term,
\begin{align}
       &\tilde{I}_{\mathrm{out}}(\omega_\out)=G(\omega_\out,\omega_0+\Omega/2)G^*(\omega_\out,\omega_0-\Omega/2),
       \label{eq:interference_term}
\end{align}
and the phase-independent term,
\begin{align}
    I_0(\w_\out)=|G(\w_\out,\w_0+\Omega/2)|^2+|G(\w_\out,\w_0-\Omega/2)|^2.
\end{align}

One can now use established methods of spectral interferometry, similar to those used in ultrafast pulse characterization techniques such as SPIDER\cite{walmsley2009characterization,davis2018experimental}, to isolate $\tilde{I}_\out(\w_\out)$ and from it extract $\phi(\w_\out,\w_\inn)$. Since $I_\out(\w_\out,\tau)$ is periodic in $\tau$, we can measure across the range $\tau \in [0, 2\pi/\Omega]$ and Fourier transform $ (\mathfrak{F}\{\cdot\})$ to obtain

\begin{align}
\mathfrak{F}\{I_\out\}(\w_\out,\bar{\omega})\sim I_0(\w_\out)\delta(\bar{\omega}) +\tilde{I}_\out(\w_\out)\delta(\bar{\omega}-\Omega)+\tilde{I}_\out^*(\w_\out)\delta(\bar{\omega}+\Omega),
\label{eq.Fourier}
\end{align}
where $\bar{\w}$ is the Fourier conjugate variable to $\tau$. We then numerically apply a filter in $\bar{\w}$-space to isolate $\tilde{I}_\out(\w_\out)$. From Eq.~\eqref{eq:interference_term}, the complex argument of $\tilde{I}_\out(\w_\out)$ is
\begin{equation}
\Delta\phi(\w_\out,\w_0)=\phi(\w_\out,\w_0+\Omega/2)-\phi(\w_\out,\w_0-\Omega/2).
\label{eq:ComplexArguement}
\end{equation} 

There are several ways to invert $\Delta\phi$ to recover the phase \cite{davis2018experimental,walmsley2009characterization}. If $\Omega$ is sufficiently small that $\phi(\w_\out,\w_\inn)$ can be considered linear between $\w_0-\Omega/2$ and $\w_0+\Omega/2$, then we can write
\begin{equation}
\Delta\phi(\w_\out,\w_0)\approx\frac{\partial\phi(\w_\out,\w_\inn)}{\partial\w_\inn}\bigg|_{_{\w_0}}\Omega.
\label{Eq.Satisfaction}
\end{equation}

If this assumption of linearity holds across wider bandwidths, then a purely two-tone signal is unnecessary and the phase can be uniquely recovered when the QFC is probed by a pulsed input (see Supplemental Document), which may be simpler in some experimental contexts.

By measuring for all values of $\omega_0$ where QFC is significant, we can recover the phase,
\begin{equation}
\phi(\w_\out,\w_\inn)=\frac{1}{\Omega}\int^{\w_\inn} \Delta\phi(\w_\out,\w_0)\dd\w_0.
\label{eq:RecoverPhase}
\end{equation}
This reconstruction of $\phi(\w_\out,\w_\inn)$ is complete up to an additive term $\chi(\w_\out)$ which is a function of $\omega_\out$ only. This unrecoverable term is equivalent to ordinary dispersive propagation after the QFC, so is of less relevance to the dynamics of the QFC itself. Substituting $G(\w_\out,\w_\inn)\rightarrow G(\w_\out,\w_\inn)e^{i\chi(\w_\out)}$ into Eq.~\eqref{efficiencyequation}, we find $\chi(\w_\out)$ cancels, so it has no bearing on process efficiency, optimal input pulse shape, or the spectral intensity of the output pulse. As with other forms of spectral shearing interferometry, the magnitude of the measured phase difference is proportional to the frequency spacing $\Omega$, so in general the measurement is more precise the larger $\Omega$.









\section{Experimental demonstration}

\begin{figure}[htbp]
    \centering
    \begin{subfigure}[b]{0.5\textwidth}
        \centering
        \includegraphics[width=\textwidth]{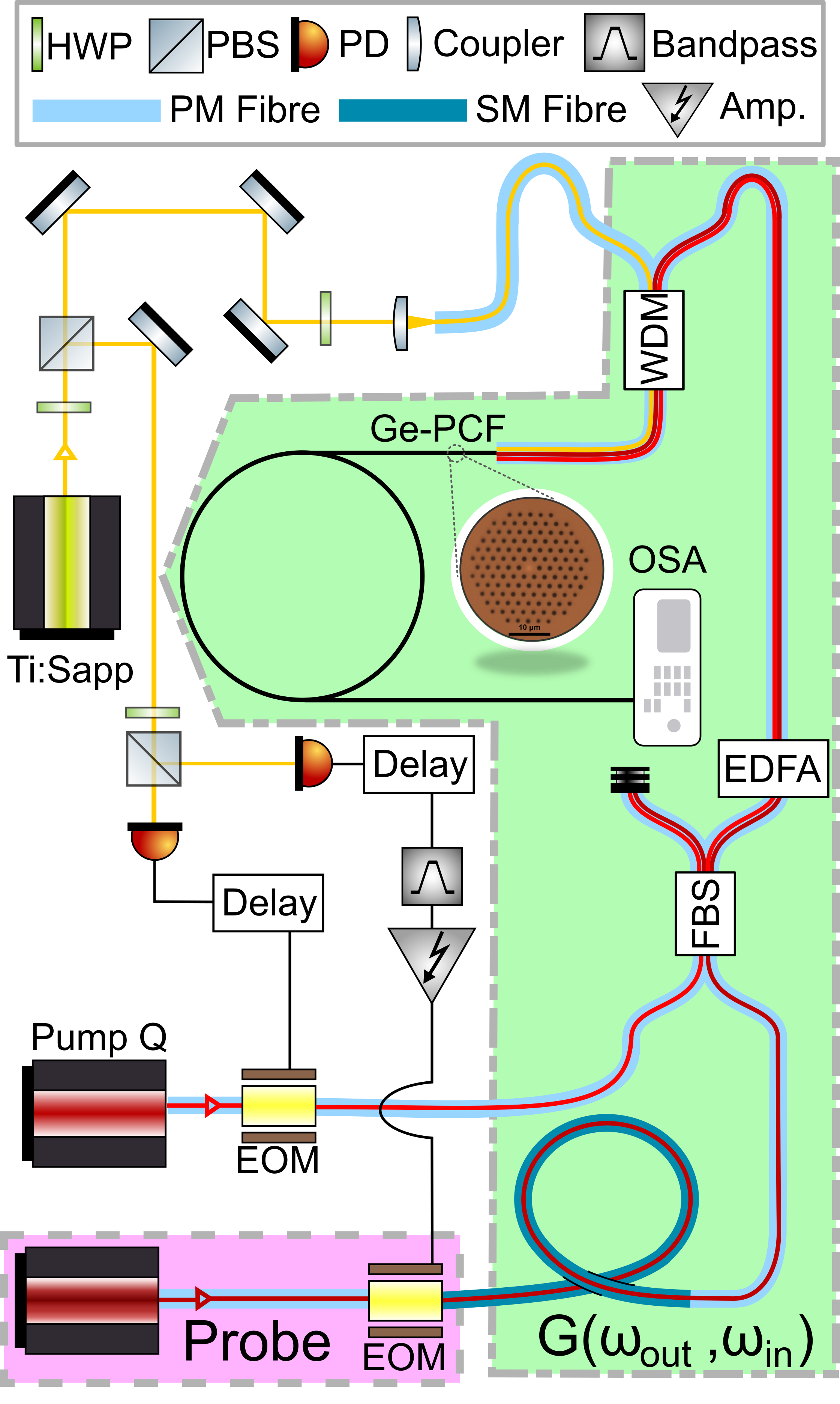}
        \caption{}
        \label{fig:ExperimentalSetup}
    \end{subfigure}
    \hfill
    \begin{subfigure}[b]{0.48\textwidth}
        \centering
        \includegraphics[width=\textwidth]{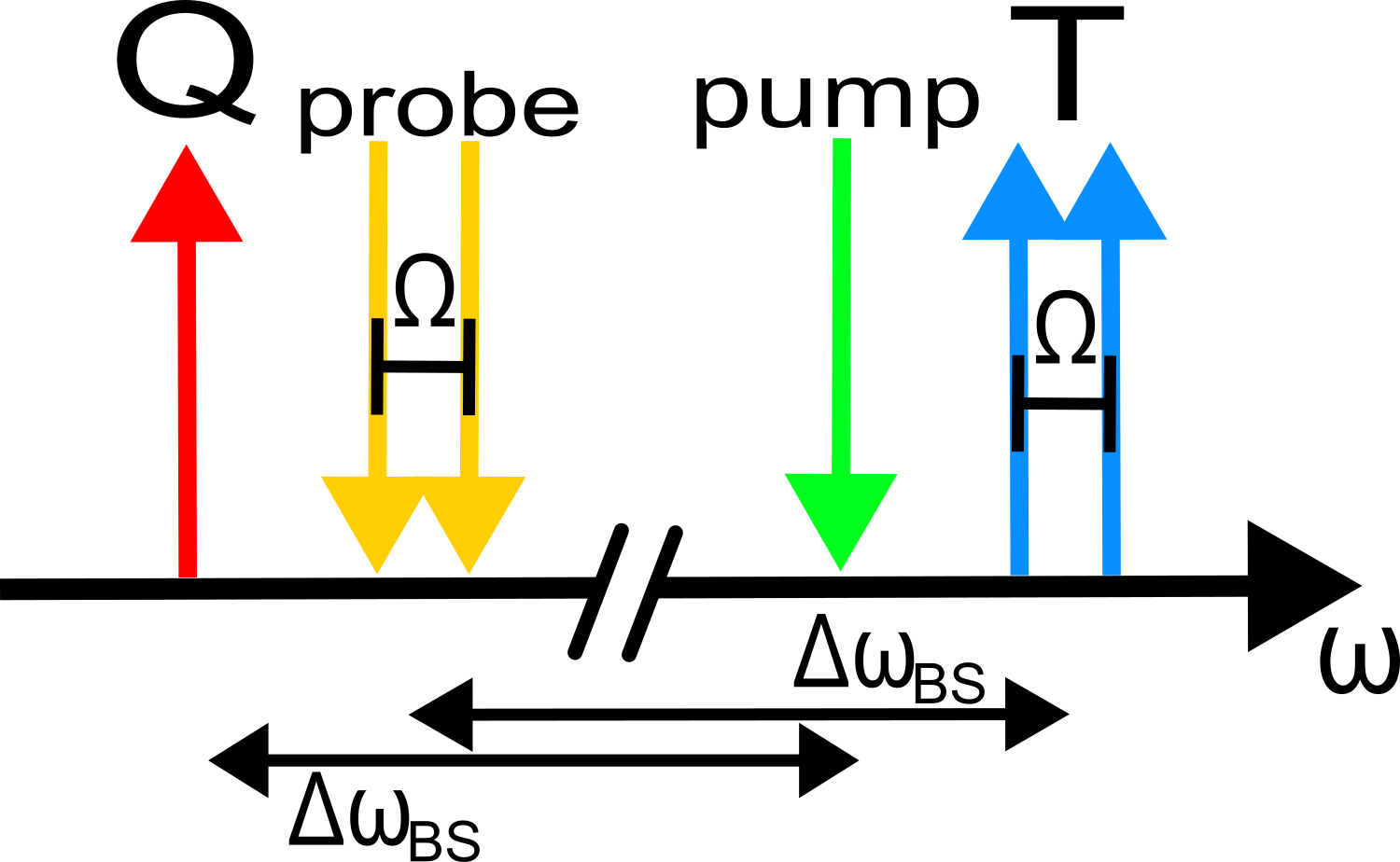}
        \caption{}
        \label{fig:Schematic}

        \includegraphics[width=\textwidth]{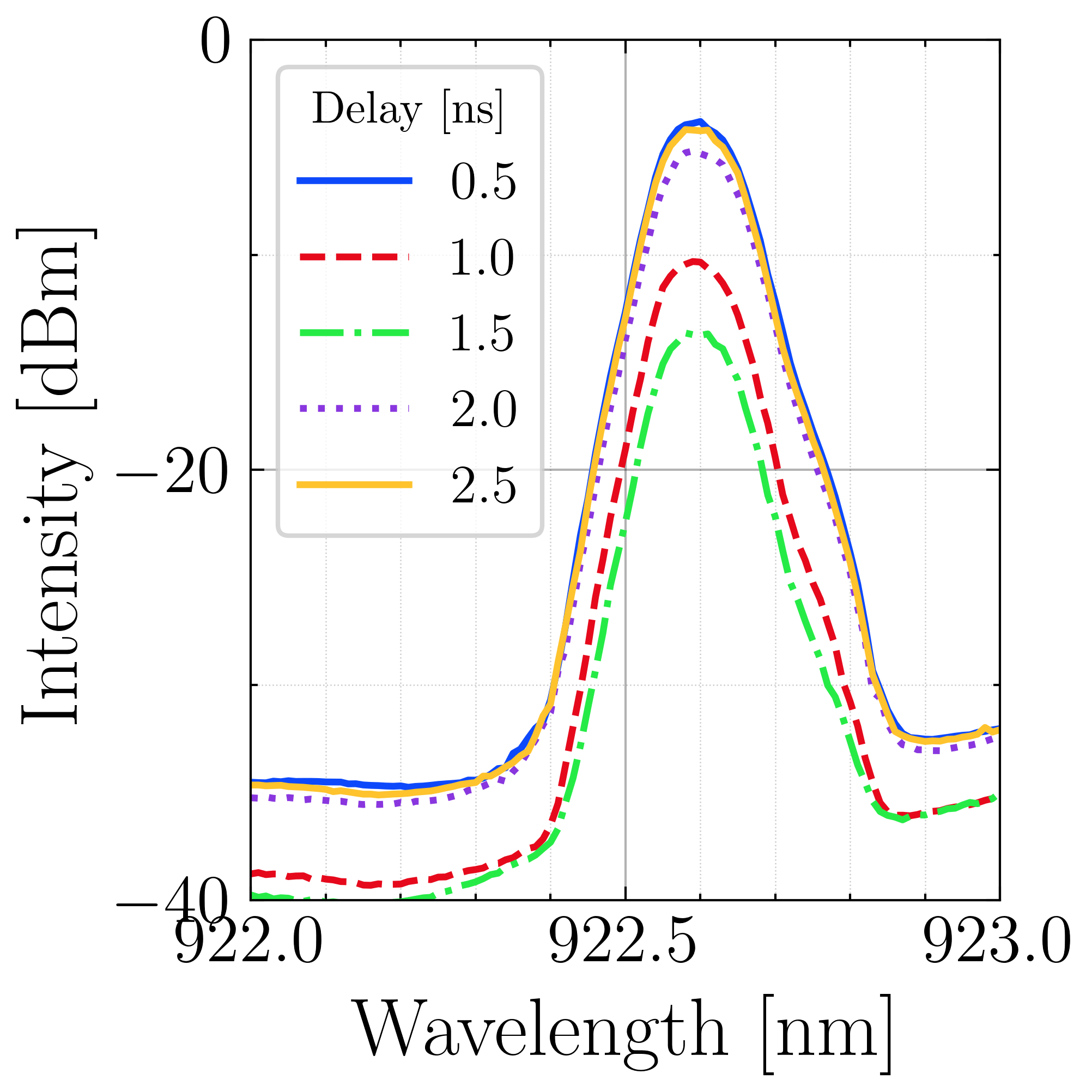}
        \caption{}
        \label{fig:DelayOSA}
    \end{subfigure}
  \caption{(a) Experimental schematic. A Ti:Sapphire laser supplies the NIR pump with a small portion picked off to create the modulation signals for the C-Band pump (Q) and probe fields, which are independently controlled by analog delay boxes. The monitoring PD drives the pulse carving for Pump Q, while the fast PD signal is bandpass-filtered and amplified to modulate the probe around \SI{560}{MHz} (highlighted pink). The Green's function $G(\omega_{\out},\omega_{\inn})$ characterizes the complete frequency conversion module (shaded green) consisting of a dispersive fiber (\SI{1.9}{km} SMF-28E+), C-Band field combination in fiber beam splitter (FBS), amplification (EDFA), wavelength division multiplexer (WDM), and the Ge-PCF (\SI{20}{m}). The output is measured by an optical spectrum analyzer (OSA). (b) Wavelength scheme for near-degenerate Bragg-scattering four-wave mixing with monochromatic pumps, illustrating frequency conversion span $\Delta\omega_{\text{BS}}$ and probe spectral shear $\Omega$. Photons are annihilated at the probe $\omega_{\text{probe}}$ and pump frequency $\omega_{\text{pump}}$, and created in the C-band pump $\omega_{\text{Q}}$ and target frequency $\omega_{\text{T}}$. (c) Variation in intensity of converted signal ($\lambda_{\text{T}}=\SI{922.63}{nm}$) as the probe ($\lambda_{\text{probe}}=\SI{1555.5}{nm}$) delay $\tau$ is swept from \SI{0.5}{ns} to \SI{2.5}{ns}.}
    \label{fig:setup}
\end{figure}

As a proof of concept for our method, we characterize a QFC module based on BS-FWM in a \SI{20}{m} segment of germanium-doped photonic crystal fiber (Ge-PCF). This device (which we reported in \cite{murphy2024tunable}) converts between light around \SI{920}{nm} (the emission window of NIR single photons from InGaAs quantum dots) and \SI{1560}{nm} (the telecom C-band). The pump scheme for this QFC is near-degenerate: the two pumps are detuned by within $\sim$10 nanometers in wavelength from the probe and target respectively (Fig. \ref{fig:Schematic}). This enables an interaction that remains efficient over relatively wide bandwidths. Here, we investigate QFC in the C-band $\rightarrow$ NIR direction. A schematic diagram of the experimental configuration is provided in Fig.~\ref{fig:ExperimentalSetup}. A Ti:Sapphire laser (SpectraPhysics Tsunami) operating at $\lambda_{\text{pump}}\approx$ \SI{923.4}{nm} with a pulse duration of \SI{35}{ps} and a repetition rate of \SI{80}{MHz} served as the master oscillator and the ultrafast pump for the BS-FWM. A small portion (\SI{5}{\%}) of the Ti:Sapphire beam was picked off and directed to a high-speed photodiode (Menlo Systems FPD610-FC), as well as a monitoring photodiode (PDA015A/M). The resulting electronic signals provide synchronized pulse carving and delay control for the C-band lasers. The remainder of the Ti:Sapphire light (average power \SI{0.210}{W}, peak power $\sim\SI{75}{W}$), was coupled into fiber and routed to a wavelength-division multiplexer (WDM) for combination with the C-Band fields before launching into the Ge-PCF.

The C-band pump (pump Q) comprises a continuous-wave (c.w.) seed laser (CoBrite DX1) with wavelength $\lambda_{Q} =$ \SI{1557.9}{nm}. The probe field was supplied by a continuously tunable c.w. C-band laser (Yenista- Tunics), which was swept across the QFC acceptance bandwidth ($\lambda_{\text{probe}} = 1551 \rightarrow \SI{1562}{nm}$). Both lasers undergo amplitude modulation in dedicated electro-optic amplitude modulators (EOM - Covega Mach 10). Pump Q is carved to nanosecond pulses which co-propagate with the picosecond Ti:Sapphire pump pulses, by directly driving the EOM with the output of the Ti:Sapphire monitoring photodiode. Carving pump Q's c.w. input into pulses increases the peak- to average-power ratio; this allows us to achieve the high peak powers required for the nonlinear interaction without saturating the average power limit of the subsequent amplifiers.

The probe is generated by driving the probe EOM with the signal from the fast photodiode, filtered using an RF bandpass filter (Mini-Circuits ZABP-650-S+). The resulting RF signal has a strong frequency component around the seventh harmonic ($\Omega=$ \SI{560}{MHz}) of the Ti:Sapphire repetition rate (Fig. \ref{fig:trace_map}). This harmonic was chosen to maximize the spectral shear, $\Omega$, within the limitation of the \SI{600}{MHz} bandwidth of the photodiode and RF amplifier. Maximizing the shear is critical because, analogous to SPIDER\cite{walmsley2009characterization,davis2018experimental}, the magnitude of the measured phase difference scales with $\Omega$. Consequently, a larger shear amplifies the phase signature against the noise floor, enhancing the signal-to-noise ratio in the spectral domain. This signal is then amplified by an RF amplifier (Mini-Circuits ZX60-83LN12+) before driving the probe EOM. By deriving the modulation frequency directly from the master oscillator, the probe signal is inherently phase-stabilized relative to the Ti:Sapphire.

To ensure temporal overlap between the Ti:Sapphire pulses and pump Q, and to facilitate temporal scanning of the probe delay $\tau$, independent electronic delays are implemented on the EOM driving signals using a multichannel analogue delay box (Ortec Model DB463) with steps of \SI{0.5}{ns}. To validate the characterization technique, we introduced a known dispersion using \SI{1.9}{km} of standard step-index optical fiber (SMF-28E+) to the probe prior to combination with the C-band pump (pump Q). 


The C-band fields (pump Q and the probe) are then combined at a fiber beam splitter and amplified by an erbium-doped fiber amplifier (EDFA- BKTel HPOA-1560-S370A). The EDFA was operated at \SI{33}{dBm} (approximately \SI{2}{W}) total output power. This operating point was chosen to maximize the nonlinear interaction strength while reducing the risk of damage to the fiber facets. The Ti:Sapphire and C-Band fields are combined at a WDM where the output is spliced to the Ge-PCF. Finally, the converted target emission is measured using an optical spectrum analyzer (OSA - Yokogawa AQ6374). 

\subsection{Phase reconstruction}

\begin{figure}[h!]
    \centering
    \begin{subfigure}[b]{0.48\linewidth}
        \centering
        \includegraphics[width=\linewidth]{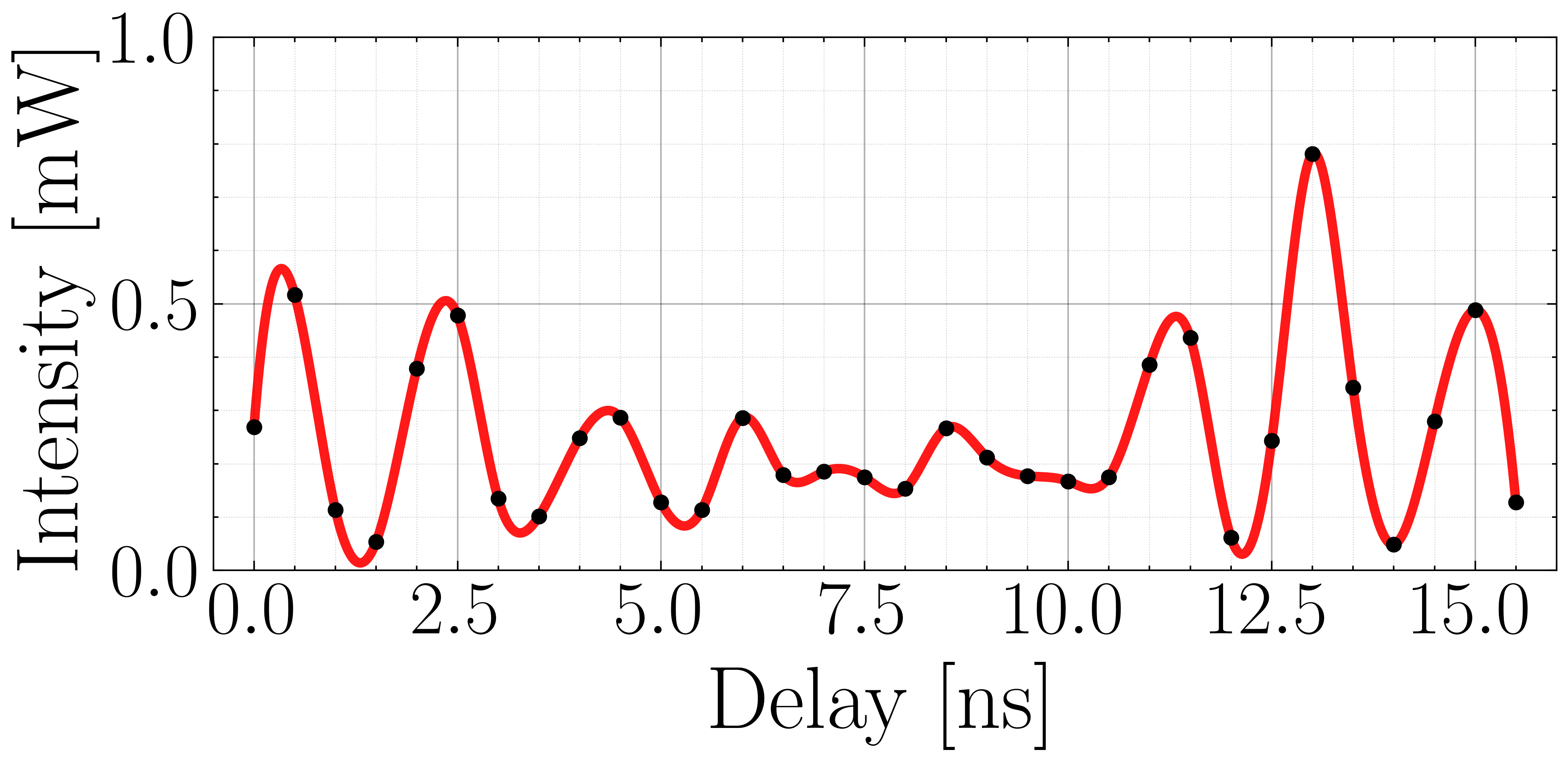}
        \caption{}
        \label{fig:trace_single}
    \end{subfigure}
    \hfill
    \begin{subfigure}[b]{0.48\linewidth}
        \centering
        \includegraphics[width=\linewidth]{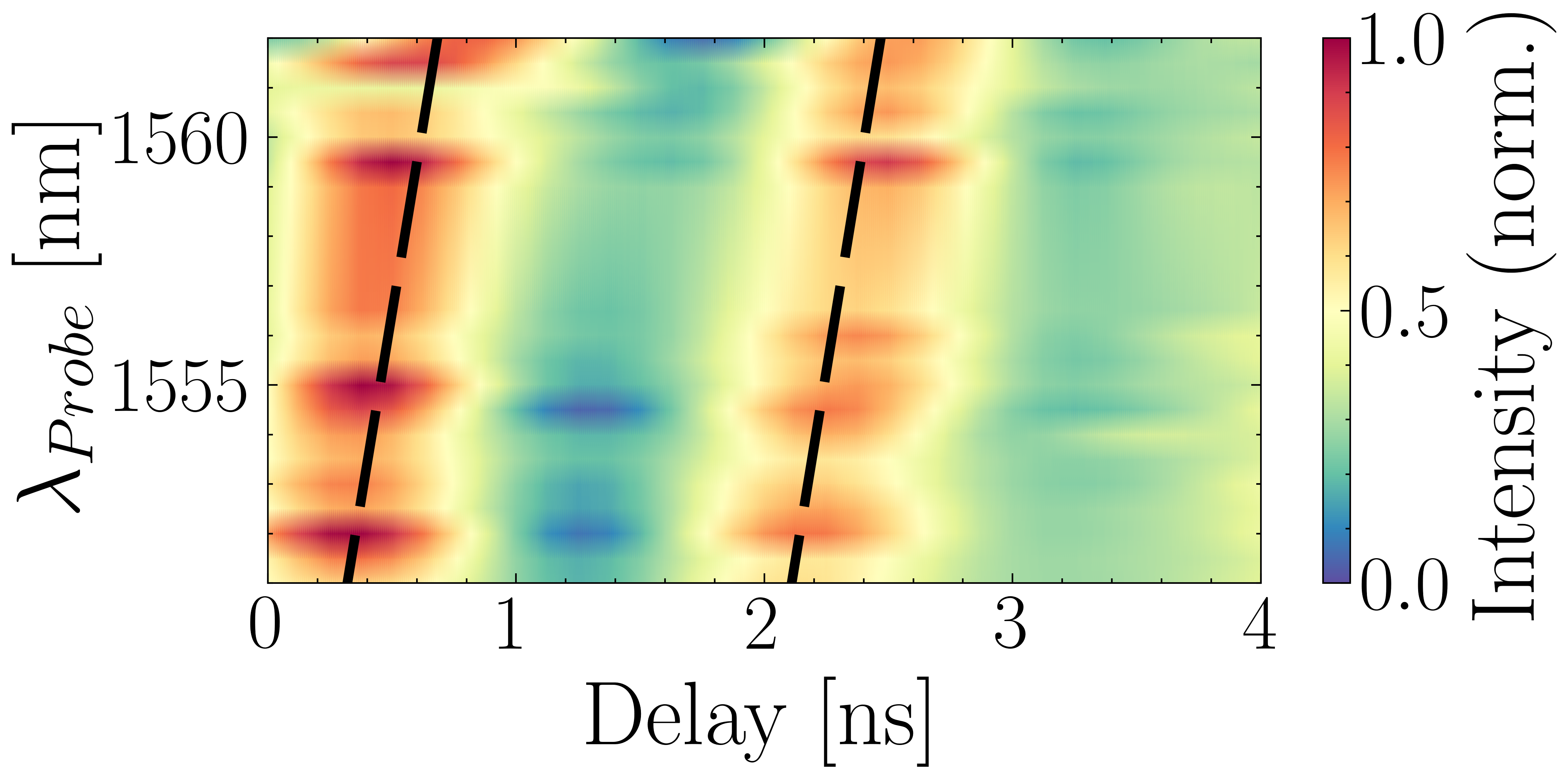}
        \caption{}
        \label{fig:trace_map}
    \end{subfigure}
    
    \medskip 
    
    \begin{subfigure}[b]{0.48\linewidth}
        \centering
        \includegraphics[width=\linewidth]{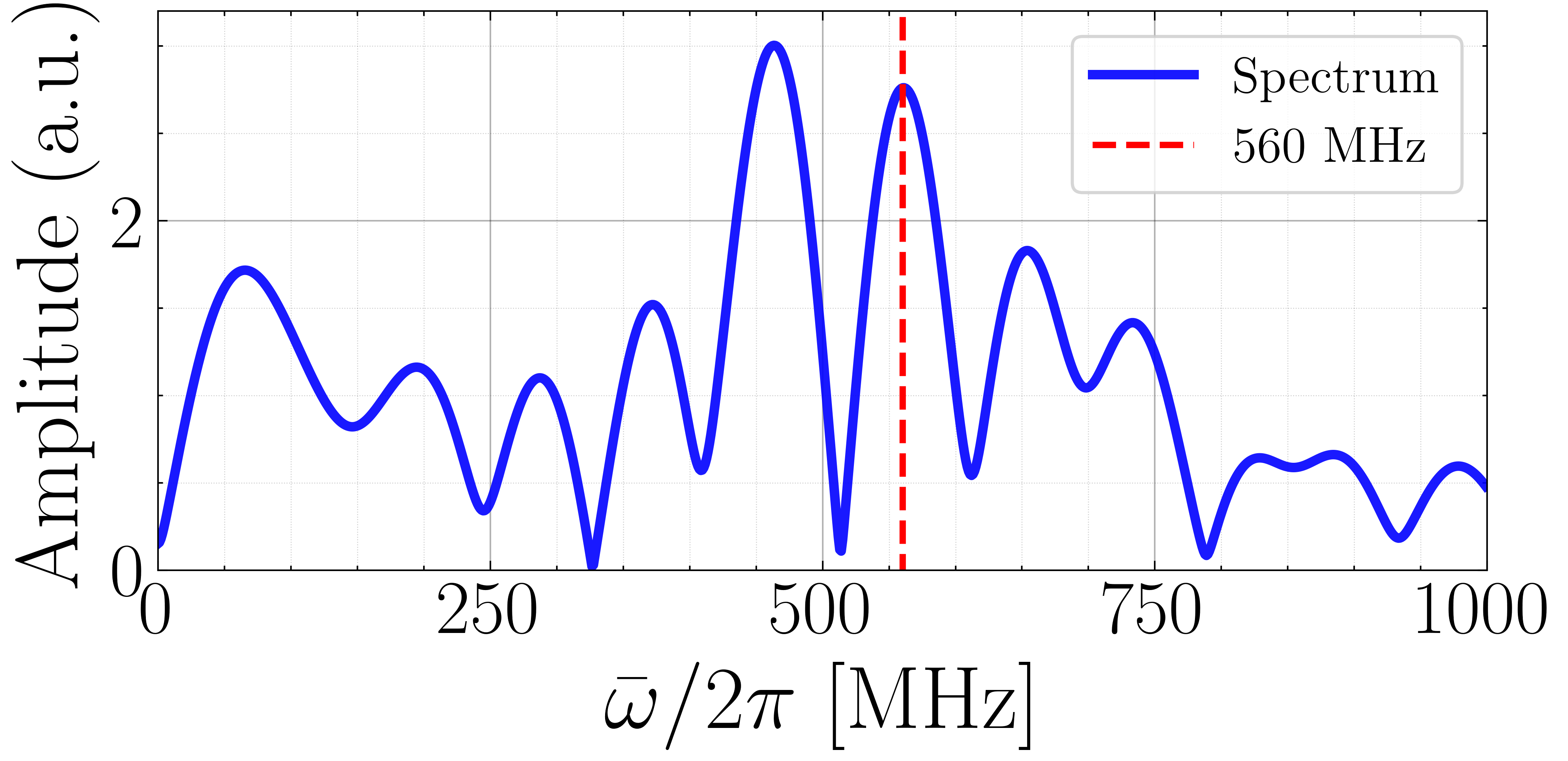}
        \caption{}
        \label{fig:fft_single}
    \end{subfigure}
    \hfill
    \begin{subfigure}[b]{0.48\linewidth}
        \centering
        \includegraphics[width=\linewidth]{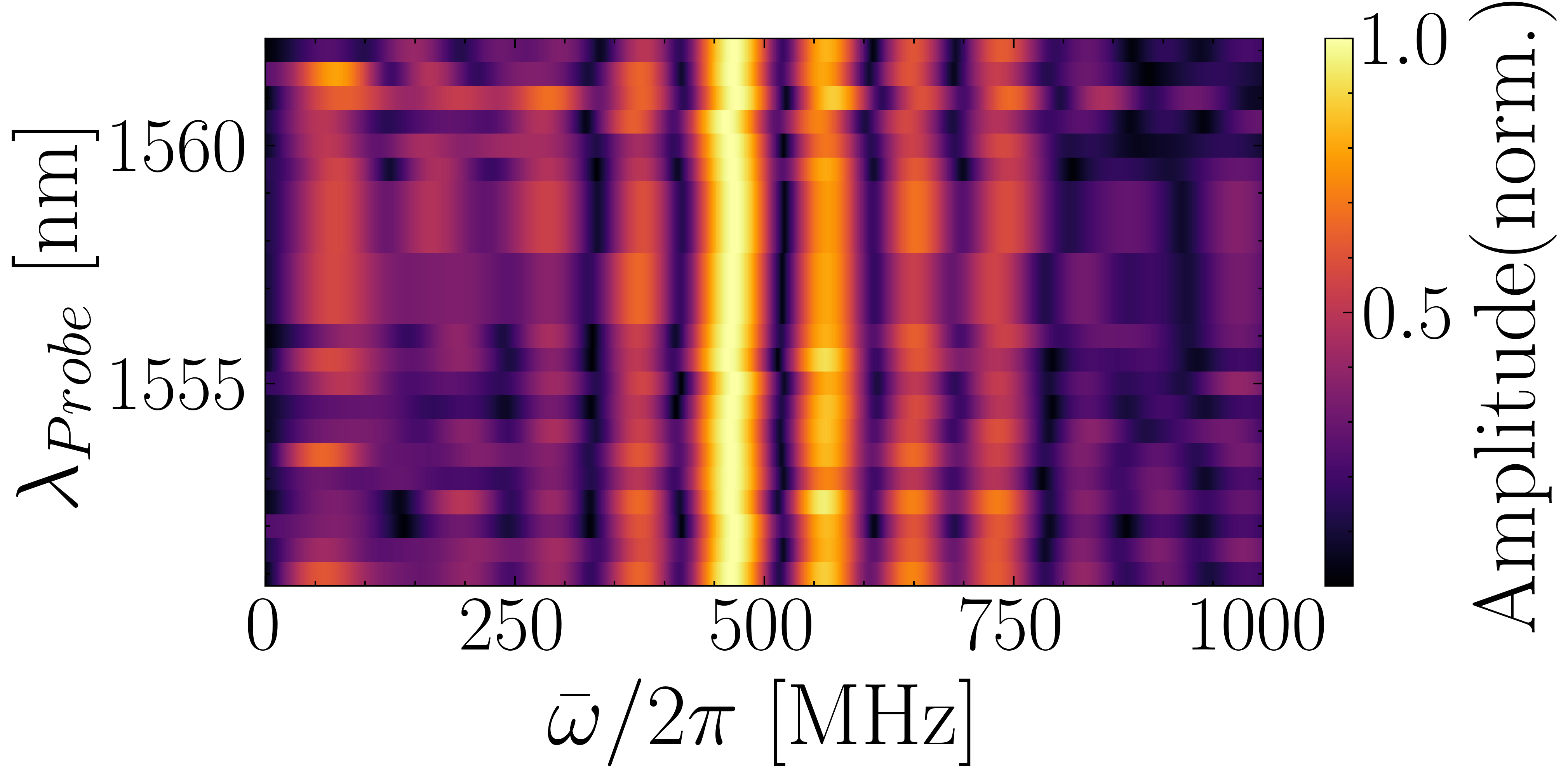}
        \caption{}
        \label{fig:fft_map}
    \end{subfigure}
    
    \caption{(a) Measured output intensity $I_{\text{out}}(\lambda_{\text{T}},\tau)$ recorded at the output wavelength for which peak conversion is achieved $\lambda_{\text{T}}=\SI{922.63}{nm}$ (corresponding to a fixed probe wavelength $\lambda_{\text{probe}}=\SI{1555.5}{nm}$), as the relative delay $\tau$ between the pumps and the probe is incremented in \SI{0.5}{ns} steps. Black dots represent experimental data, while the solid red line denotes the interpolation used for resampling. (b) Spectrogram of the target emission intensity across the full probe tuning range. The data is displayed over 0-\SI{4}{ns} temporal window to highlight the diagonal fringes resulting from the group velocity dispersion, emphasized by two parallel dashed lines separated by $1/\SI{560}{MHz}$. (c) Fourier transform of the time domain trace shown in (a) into the conjugate frequency space, $\bar{\omega}$, revealing a peak at the modulation frequency $\Omega=\SI{560}{MHz}$ (the adjacent sixth harmonic is also prominent). (d) The magnitude of the spectrally resolved Fourier transform, $|\tilde{I}_{\text{out}}(\omega_{\text{out}},\bar{\w})|$, for all probe wavelengths, from which the phase information is extracted.}
    \label{fig:methodology_grid}
\end{figure}

The reconstruction of the Green's function relies on isolating the phase-dependent interference term, $\tilde{I}_{\text{out}}(\omega_{\text{out}})$, from the intensity beats generated by the bichromatic probe. An example of this interference is illustrated in Fig.~\ref{fig:DelayOSA} by examining the system response at a fixed probe wavelength of $\lambda_{\text{probe}}=\SI{1555.5}{nm}$ corresponding to the converted wavelength $\lambda_{\text{T}}=\SI{922.63}{nm}$. By sweeping the electronic delay $\tau$ between the dual pump pulse train and the probe modulation, a characteristic oscillation in conversion efficiency is observed; this variation constitutes the time domain signature of the spectral interference required for characterization.

The full temporal evolution of this signal was recorded by the scanning of $\tau$, in \SI{0.5}{ns} steps, as shown in Fig. \ref{fig:trace_single}. To maximize the signal-to-noise ratio, three consecutive spectral measurements were taken and averaged at each delay step using the OSA, which acquired spectra between \SI{920}{nm} and \SI{930}{nm} with a resolution of \SI{0.05}{nm}.

This measurement was repeated for a range of values of $\lambda_{\text{probe}}$ spanning the acceptance window of the QFC. The resulting time-frequency interference fringes are shown in Fig \ref{fig:trace_map}. In this plot, the data is restricted to a 0-\SI{4}{ns} temporal window to emphasize the diagonal slant of the fringes. This linear temporal shift provides a direct visualization of the variation in group delay across the spectrum, reflecting the dispersive propagation inside the QFC module.

To reconstruct the complex Green's function, the multidimensional dataset was processed using a vectorized approach. For each output wavelength, the recorded time domain intensity traces were resampled using a Fourier domain interpolation to provide a high-resolution, uniform temporal grid, $\delta\tau$. The spectrally resolved traces were then transformed into the Fourier conjugate space ($\bar{\omega}$) of the delay $\tau$ as defined in Eq.~\eqref{eq.Fourier}. This transformation, shown in Fig.~\ref{fig:fft_single}, reveals the spectral content of the modulation beat and isolates the interference terms at $\pm\Omega$. The complex coefficient, $\tilde{I}_{\text{out}}(\omega_{\text{out}},\bar{\w})$, was extracted to serve as the signal; the full spectrally resolved Fourier transform across all probe wavelengths is shown in Fig.~\ref{fig:fft_map}. 

The argument of the extracted complex coefficient provides the relative phase shift $\Delta\phi$ between the probe frequency components as specified in Eq.~\eqref{eq:ComplexArguement}. This relation allows the probe input group delay to be recovered as $\tau_g =\frac{\partial\phi}{\partial\omega_\inn} \approx\Delta\phi/\Omega$. The complete spectral phase $\phi(\omega_{\text{out}},\omega_{\text{in}})$ can then be reconstructed by integrating the phase differences according to Eq.~\eqref{eq:RecoverPhase}.

\subsection{Results and discussion}

\begin{figure}[h!]
  \begin{subfigure}[t]{.48\textwidth}
    \centering
    \includegraphics[width=\linewidth]{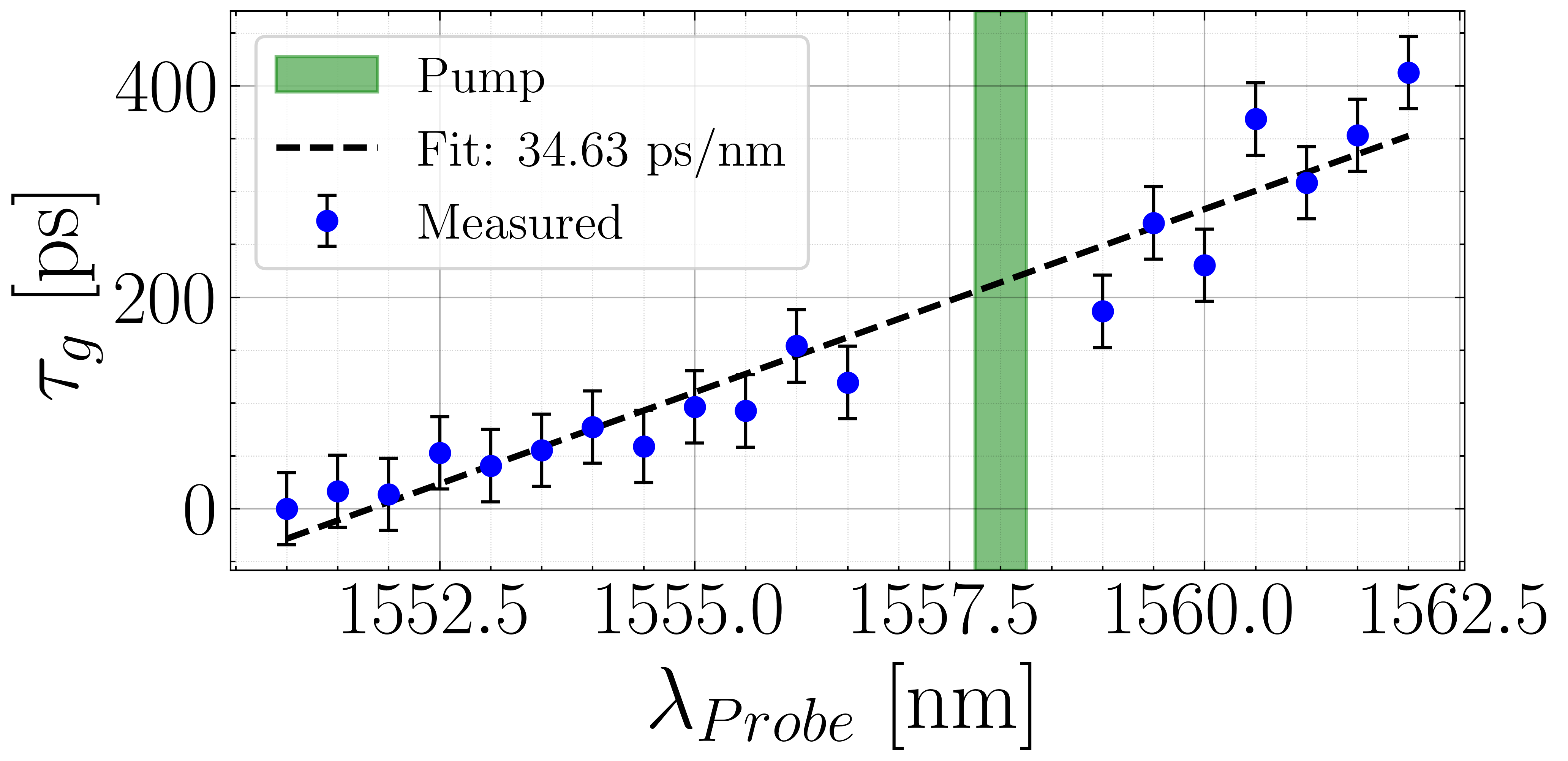}
    \caption{}
    \label{fig:SubResult_a}
  \end{subfigure}
  \hfill
  \begin{subfigure}[t]{.48\textwidth}
    \centering
    \includegraphics[width=\linewidth]{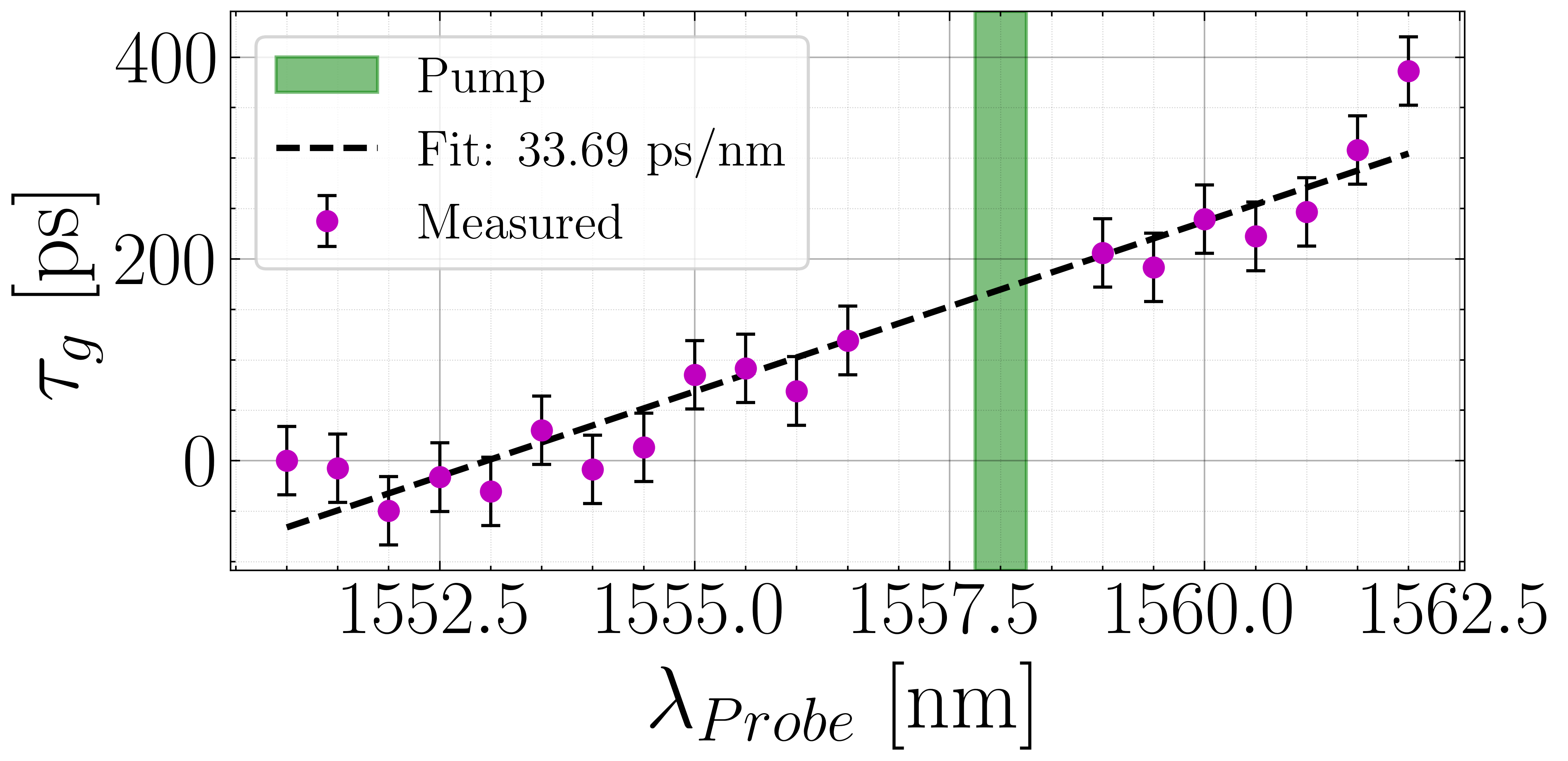}
    \caption{}
    \label{fig:SubResult_b}

  \end{subfigure}

  \medskip

    \begin{subfigure}[t]{.48\textwidth}
    \centering
    \includegraphics[width=\linewidth]{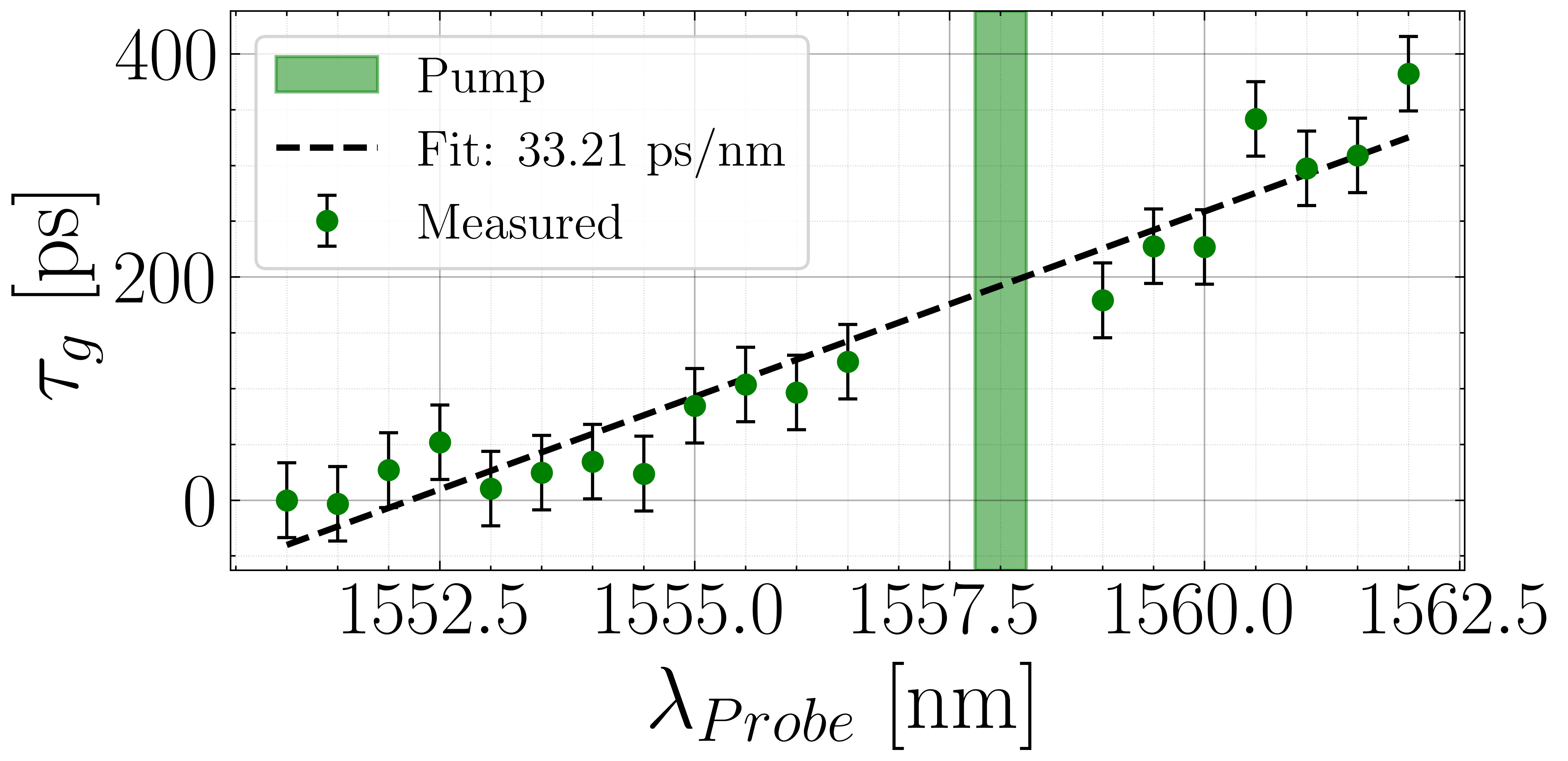}
    \caption{}
    \label{fig:SubResult_c}
  \end{subfigure}
  \hfill
  \begin{subfigure}[t]{.48\textwidth}
    \centering
    \includegraphics[width=\linewidth]{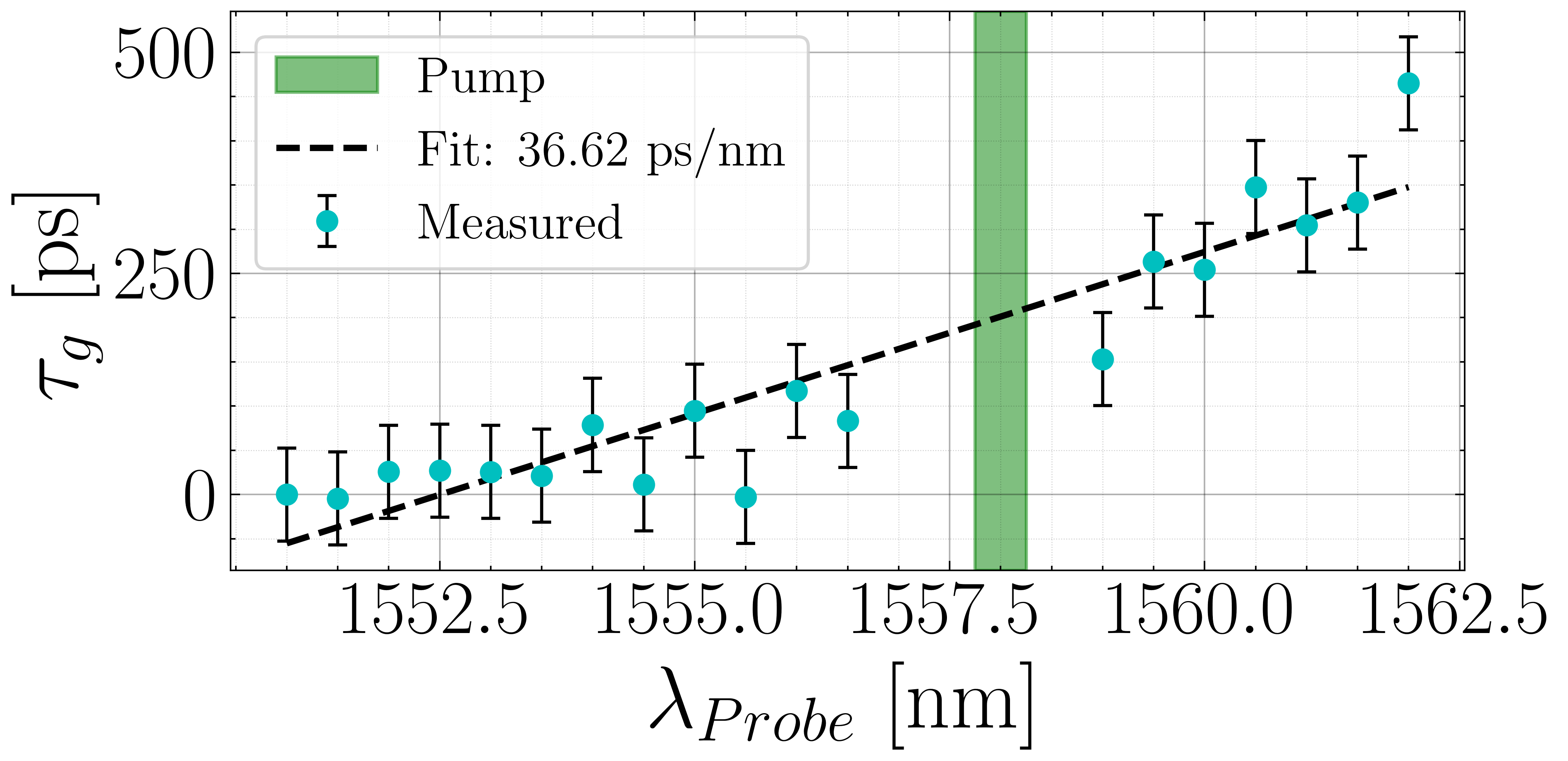}
    \caption{}
    \label{fig:SubResult_d}
  \end{subfigure}
  \caption{(a) Measured relative group delay, $\tau_{g}$, as a function of probe wavelength for the primary shear frequency $\Omega=\SI{560}{MHz}$. The dashed line represents the linear fit yielding a dispersion slope of \SI{34.63}{ps/nm} with a mean $1\sigma$ uncertainty of $\SI{\pm34.2}{ps}$. (b)-(d) Group delay retrieval for adjacent harmonics of the laser repetition rate used for validation: (b) \SI{480}{MHz} ($\SI{33.69}{ps/nm}$, $\SI{\pm 33.9}{ps}$); (c) \SI{640}{MHz} ($\SI{33.21}{ps/nm}$ , $\SI{\pm 33.4}{ps}$ ); and (d) \SI{720}{MHz} ($\SI{36.62}{ps/nm}$ , $\SI{\pm 52.6}{ps}$). All error bars represent $1\sigma$ r.m.s. deviation from the linear fit.}
  \label{fig:SubResult}
\end{figure}


The results for the shear frequency $\Omega =\SI{560}{MHz}$ are presented in Fig. \ref{fig:SubResult_a}. A linear fit to the measured group delay yields a constant dispersion slope of \SI{34.63}{ps/nm}. The error bars plotted across Fig. \ref{fig:SubResult} represent the residual errors of the measurement, calculated as the r.m.s. individually for each dataset. This yields a $1\sigma$ uncertainty of $\SI{\pm34.2}{ps}$ for the \SI{560}{MHz} data. 

We validated the accuracy of the Green's function reconstruction by recovering the predicted dispersive signature of the optical path. Based on a standard dispersion of $D \approx \SI{18.0}{ps/(nm\cdot km)}$ at \SI{1550}{nm}, the introduced dispersion is expected to induce a delay slope of \SI{34.2}{ps/nm}. The total length of the active and auxiliary fiber components (<\SI{30}{m}) is negligible by comparison. The measured dependence of the phase on $\omega_\out$ reflects the total system Green's function response (highlighted green in Fig. \ref{fig:ExperimentalSetup}), confirming that the technique correctly recovers the spectral phase.

\begin{figure}[htbp]
    \begin{subfigure}[b]{0.48\textwidth}
        \centering
        \includegraphics[width=\linewidth]{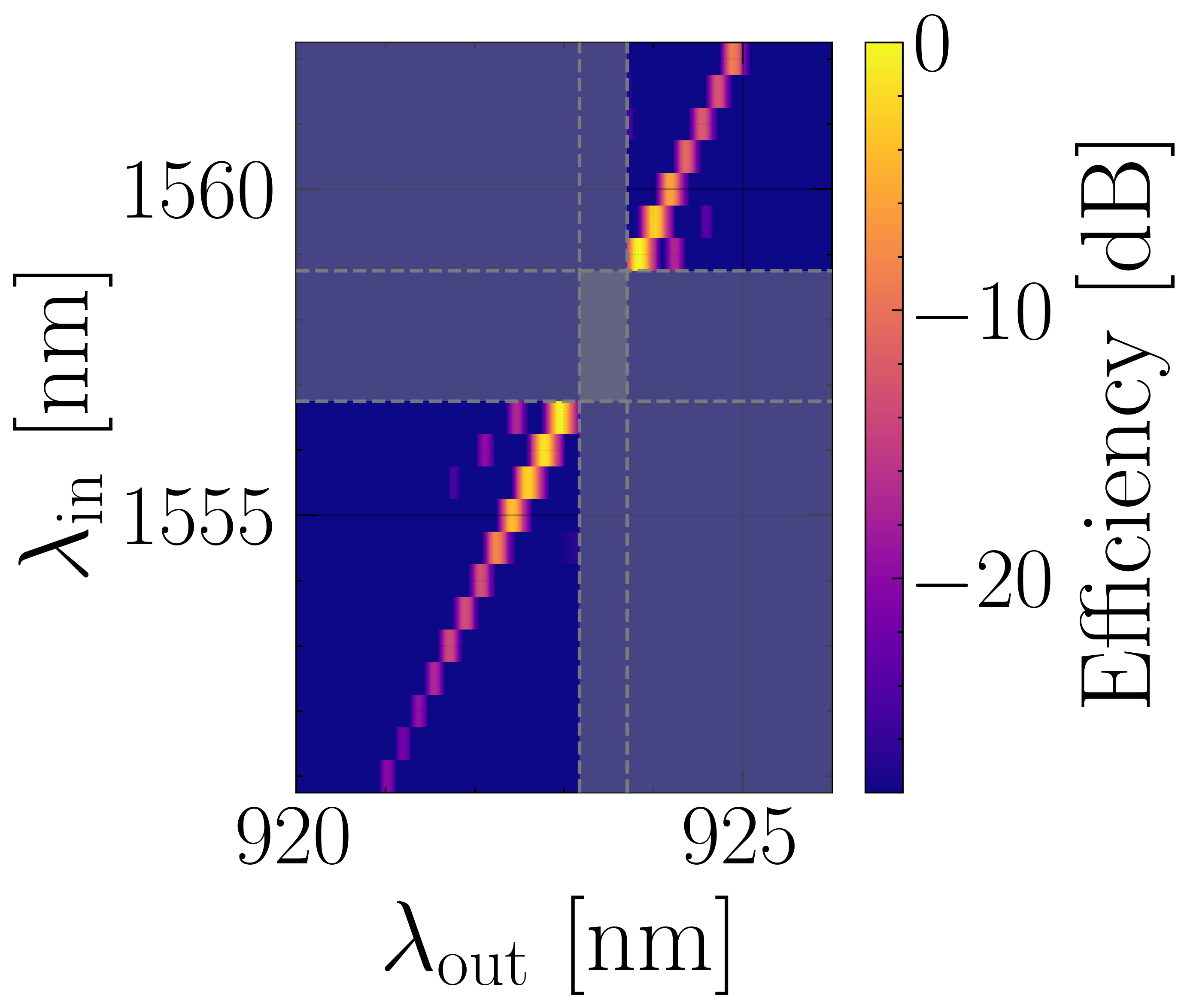}
        \caption{}
        \label{fig:Master_a}
    \end{subfigure}
    \hfill 
    \begin{subfigure}[b]{0.48\textwidth}
        \centering
        \includegraphics[width=\linewidth]{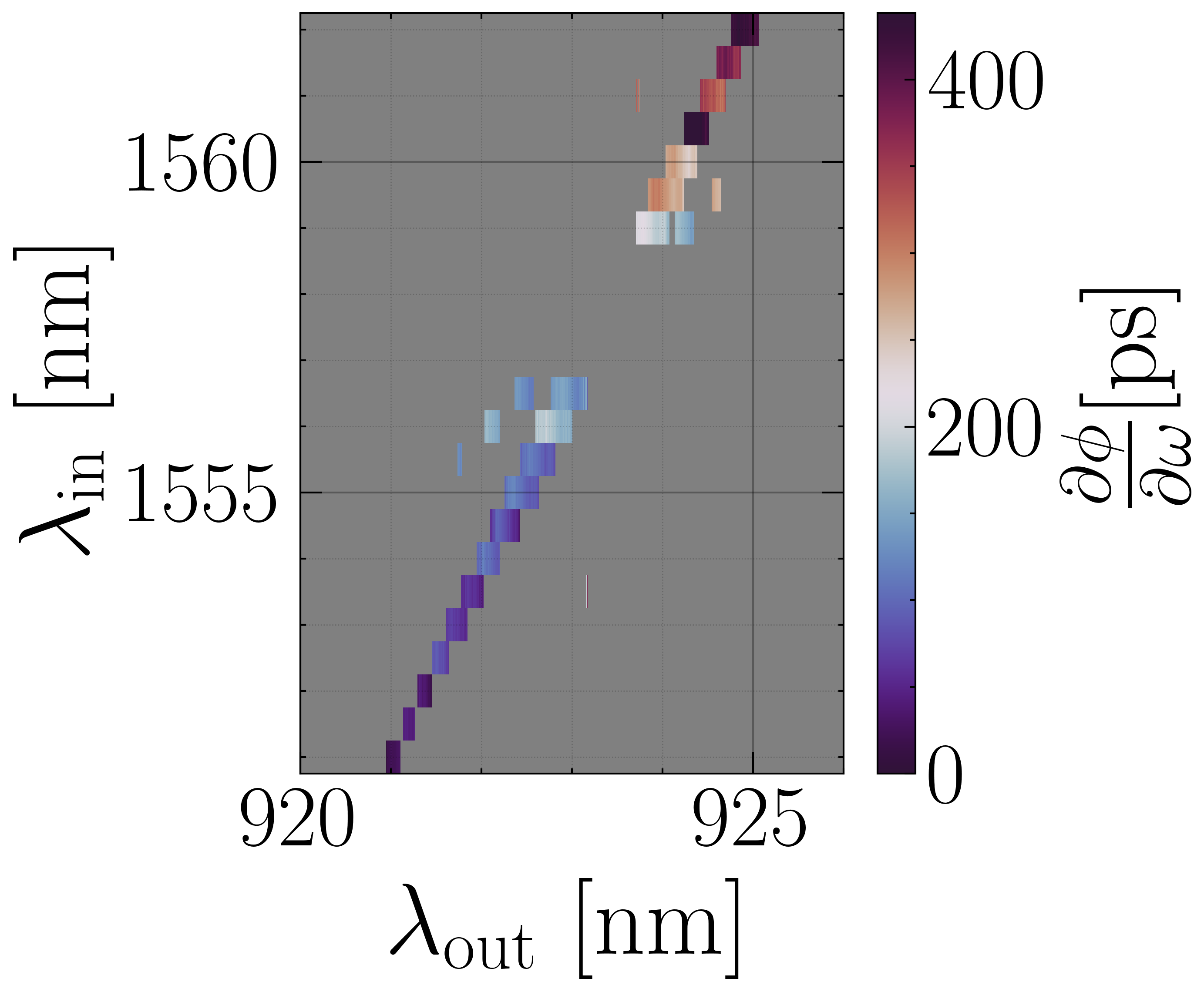}
        \caption{}
        \label{fig:Master_b}
    \end{subfigure}

\caption{Retrieval of the complex Green's function. (a) Reconstructed spectral intensity, $|G(\omega_{\text{out}},\omega_{\text{in}})|^2$. (b) Two dimensional map of the relative group delay $\tau_g = \frac{\partial\phi}{\partial\omega}$.}  
    \label{fig:MasterResult}
\end{figure}

We further verified the robustness of the technique by analyzing subsidiary probe harmonics. The recovered dispersion slopes at \SI{480}{MHz}, \SI{640}{MHz} and \SI{720}{MHz}, Fig. \ref{fig:SubResult_b} to \ref{fig:SubResult_d}, gave \SI{33.69}{ps/nm} ($\sigma_{\tau} \approx \SI{33.9}{ps}$), \SI{33.21}{ps/nm} ($\sigma_{\tau} \approx \SI{33.4}{ps}$) and \SI{36.62}{ps/nm} ($\sigma_{\tau} \approx \SI{52.6}{ps}$) respectively. This agreement confirms that the shear frequencies employed are sufficiently small to satisfy the linear phase approximation required by Eq.~\eqref{Eq.Satisfaction}. The retrieved values for the amplitude and phase derivative of $G(\w_\out,\w_\inn)$ are summarized in Fig. \ref{fig:MasterResult}.

Figure \ref{fig:Master_a} displays the reconstructed spectral intensity, $|G(\omega_{\text{out}}, \omega_{\text{in}})|^2$, normalized to the maximum value. Regions containing the pump fields and parasitic FWM processes, which are excluded from the linear Green's function formalism, Eq. \ref{Gdef}, are masked in gray. The corresponding phase information is presented in Fig. \ref{fig:Master_b} as a full two-dimensional map of the relative group delay, $\tau_g = \frac{\partial\phi}{\partial\omega} \approx\frac{\Delta\phi}{\Omega}$. By retrieving this phase information alongside the spectral amplitude, we demonstrate the capability to fully characterize the conversion process, providing the essential data required to synthesize optimal input states for high-efficiency quantum networking.

While chromatic dispersive delays of hundreds of picoseconds are clearly resolved, the picosecond-scale phase variations arising from meter-scale internal conversion dynamics of the Ge-PCF currently remain below our experimental resolution. This was primarily limited by the relatively small value of $\Omega$. Electronic distortion of the modulation signal introduced by the analog delay box (used to scan $\tau$ over correspondingly large delays of order $\pi/\Omega$) was also a significant issue. This limitation was technical rather than fundamental, and could be overcome by using a digital delay generator. Related experiments in quantum pulse characterization have successfully employed arbitrary waveform generators or parametric dielectric resonant oscillators to generate phase-stabilized EOM drive signals at frequencies up to \SI{40}{GHz} (the 500th harmonic of the \SI{80}{MHz} repetition rate) \cite{wright2017spectral}, which can be delayed without distortion. Implementing such a high-frequency driver would increase the measurement sensitivity by a factor of more than 70, bringing femtosecond-scale chromatic delays within reach.

\section{Conclusion and outlook}

We have proposed and experimentally demonstrated a technique for the complete, phase-sensitive characterization of arbitrary unitary spectral-temporal transformations in QFC. Our numerical analysis highlights how the form of the Green's function spectral phase arises from QFC dynamics, and its characterization is critical for maximizing conversion efficiency by optimizing the input mode (or pump pulse to tailor the QFC process for maximum conversion of a specific desired input pulse shape). We have shown that this complex Green's function $G(\omega_{\text{out}} \omega_{\text{in}})$ can be fully recovered up to a phase term $e^{i\chi(\w_\out)}$ that depends on the output frequency only. 

We validated this proposal by experimentally reconstructing the complex Green's function, $G(\omega_{out}, \omega_{in})$, for a composite QFC module consisting of dispersive propagation through \SI{1.9}{km} SMF-28E+ followed by BS-FWM in a \SI{20}{m} segment of Ge-PCF. The retrieved spectral phase encoded the internal dynamics of the frequency converter, revealing the presence of the dispersive propagation before the region of active conversion. The extracted dispersion slope of \SI{34.63}{ps/nm} aligns with the theoretical prediction of \SI{34.2}{ps/nm}, confirming the interferometric phase retrieval derived in Section \ref{theory}. 



While our experiment realizes a proof of principle, it was limited by the frequency shear of the bichromatic probe and RF signal distortion incurred by the correspondingly large temporal delays required. We have described how available methodology can improve on our demonstration by some orders of magnitude, enabling detailed reconstruction of QFC dynamics down to femtosecond timescales and precise optimization of QFC processes for conversion efficiency.

As quantum technologies evolve, seamlessly interfacing spectrally distinct memories and sources demands precise spectral-temporal mode matching. The applicability of our method extends far beyond the Ge-PCF platform demonstrated here. Relying solely on the linearity of the conversion process, it is universally compatible with all photon-based QFC modalities, encompassing both $\chi^{(2)}$ and $\chi^{(3)}$ interactions. By delivering a scalable and truly platform-agnostic solution for full process tomography, this tool provides the essential diagnostic framework required to engineer the active quantum devices that will underpin future long-distance quantum communications.

\section{Acknowledgements}
The authors gratefully acknowledge the advanced fiber fabrication research facility at the University of Bath \cite{BathResearchFacilities}, with thanks to Steven Renshaw for his technical support. We are also grateful to Prof William Wadsworth and Prof Tim Birks for use of equipment.

This work was funded by the UK National Quantum Technologies Programme through EPSRC grants PHOCIS (EP/W028336/1) and QCI3 (EP/Z53318X/1).

\bibliographystyle{unsrt}
\bibliography{bibliography}

\end{document}